\documentclass[sigconf]{acmart}
    
\usepackage{multirow}
\usepackage{balance}
\usepackage{caption}
\usepackage{subcaption}
\usepackage{algorithm}
 
\usepackage{amssymb}
\usepackage{bm}

\copyrightyear{2021}
\acmYear{2021}
\setcopyright{acmcopyright}\acmConference[KDD '21]{Proceedings of the 27th ACM SIGKDD Conference on Knowledge Discovery and Data Mining}{August 14--18, 2021}{Virtual Event, Singapore}
\acmBooktitle{Proceedings of the 27th ACM SIGKDD Conference on Knowledge Discovery and Data Mining (KDD '21), August 14--18, 2021, Virtual Event, Singapore}
\acmPrice{15.00}
\acmDOI{10.1145/3447548.3467140}
\acmISBN{978-1-4503-8332-5/21/08}


\begin{document}
\fancyhead{}

\title{Dual Attentive Sequential Learning for Cross-Domain Click-Through Rate Prediction}

\author{Pan Li}
\affiliation{%
  \institution{New York University}
  \streetaddress{44 West 4th Street}
  \city{New York}
  \country{USA}}
\email{pli2@stern.nyu.edu}

\author{Zhichao Jiang}
\affiliation{%
  \institution{Alibaba Youku Cognitive and Intelligent Lab}
  \city{Beijing}
  \country{China}}
\email{zhichao.jzc@alibaba-inc.com}

\author{Maofei Que}
\affiliation{%
  \institution{Alibaba Youku Cognitive and Intelligent Lab}
  \city{Beijing}
  \country{China}}
\email{maofei.qmf@alibaba-inc.com}

\author{Yao Hu}
\affiliation{%
  \institution{Alibaba Youku Cognitive and Intelligent Lab}
  \city{Beijing}
  \country{China}}
\email{yaoohu@alibaba-inc.com}

\author{Alexander Tuzhilin}
\affiliation{%
  \institution{New York University}
  \streetaddress{44 West 4th Street}
  \city{New York}
  \country{USA}}
\email{atuzhili@stern.nyu.edu}

\begin{abstract}
Cross domain recommender system constitutes a powerful method to tackle the cold-start and sparsity problem by aggregating and transferring user preferences across multiple category domains. Therefore, it has great potential to improve click-through-rate prediction performance in online commerce platforms having many domains of products. While several cross domain sequential recommendation models have been proposed to leverage information from a source domain to improve CTR predictions in a target domain, they did not take into account bidirectional latent relations of user preferences across source-target domain pairs. As such, they cannot provide enhanced cross-domain CTR predictions for both domains simultaneously. In this paper, we propose a novel approach to cross-domain sequential recommendations based on the dual learning mechanism that simultaneously transfers information between two related domains in an iterative manner until the learning process stabilizes. In particular, the proposed Dual Attentive Sequential Learning (DASL) model consists of two novel components Dual Embedding and Dual Attention, which jointly establish the two-stage learning process: we first construct dual latent embeddings that extract user preferences in both domains simultaneously, and subsequently provide cross-domain recommendations by matching the extracted latent embeddings with candidate items through dual-attention learning mechanism. We conduct extensive offline experiments on three real-world datasets to demonstrate the superiority of our proposed model, which significantly and consistently outperforms several state-of-the-art baselines across all experimental settings. We also conduct an online A/B test at a major video streaming platform Alibaba-Youku, where our proposed model significantly improves business performance over the latest production system in the company.
\end{abstract}



\maketitle

\section{Introduction}
Recommender systems have become increasingly important for online marketplaces to achieve their business success. Among common recommendation tasks, Click-Through-Rate (CTR) prediction has generated significant research interest because it is a practically important business performance measure influencing user behavior \cite{zhou2018deep,lian2018xdeepfm} that is also closely related to the revenue generating performance metrics used in online platforms. In particular, researchers have proposed to use sequential recommendation models \cite{sutskever2014sequence,wang2019sequential} to automatically extract latent user and item representations and their high-order interactions to achieve better CTR prediction performance, including GRU \cite{hidasi2015session}, LSTM \cite{zhu2017next}, and BERT \cite{sun2019bert4rec}. However, these methods also suffer from the cold start and data sparsity problems \cite{schein2002methods,adomavicius2005toward} commonly encountered in recommendation tasks, which can severely affect the performance metrics since we only have access to a small proportion of past consumption records. 

To address these problems, researchers have proposed to use cross domain recommender systems \cite{cantador2015cross} to transfer and aggregate user preference across different domains \cite{pan2010survey}. For example, cross domain sequential recommendation models \cite{ma2019pi,ouyang2020minet} leverages auxiliary user information from a source domain to improve the CTR prediction performance of a target domain. However, existing cross domain sequential recommendation models primarily focus on the unidirectional transfer of user preferences from the source domain to the target domain without taking into account the dual nature of providing recommendations in source-target domain pairs. While achieving satisfying CTR prediction performance in the target domain, these models might not achieve the optimal CTR prediction performance in the original source domain, which is unfortunate as industrial platforms typically have many category domains, and it is important to improve the recommendation performance in all domains simultaneously.

Meanwhile, it is more desirable to also transfer user preferences in the other direction, i.e., from the target domain back to the source domain, since improving recommendation performance in one domain could also lead to the improvements of recommendation performance in the other domain \cite{li2020ddtcdr}, especially for CTR predictions as users may simultaneously seek purchases of items from multiple categories such as buying electronics for Christmas gifts and candles for Christmas decorations. If we can provide better CTR predictions for the electronics domain, we can use those predictions to improve our recommendations for the decorations domain, and vice versa.

Therefore in this paper, we propose to apply the dual learning mechanism to sequantial learning models to bidirectionally transfers user preferences across different domains simultaneously for cross-domain click-through rate prediction task. Our proposed dual learning-based model consists of the following two novel components:

\begin{itemize}
\item \textbf{Dual Embedding}. As user purchasing actions in a certain domain could be affected by their exposures to previous recommendations in other domains, we propose a novel \emph{dual embedding} component that unifies the learning process of user representations. In particular, we adopt the idea of metric learning \cite{kulis2012metric} and map the latent user and item representations in different domains into the shared latent space, where we aim to minimize the metric distances between different representations. By doing so, we could not only bring two item embeddings closer to each other if they are purchases by the same user, but also bring two user embeddings closer to each other through the similar items consumed across the two domains. Therefore, we progressively improve the modeling of user preferences for all domains through the iterative learning process in the Dual Embedding component.

\item \textbf{Dual Attention}. As the transaction records for sequential recommendations often contain many items irrelevant to the next choice, attention mechanism \cite{shaw2018self,zhou2018deep} is often utilized to weight the observed items with different relevance to build an attentive context that outputs the proper next item with a high probability. In this paper, we propose a novel \emph{dual attention} component to estimate relative importance of historical items in \emph{both} domains simultaneously. Unlike classical attention-based recommendation models that only takes into account user behavior in one single domain, the Dual Attention component incorporates user behaviors in multiple different domains to obtain the associated attention values.
\end{itemize}

\noindent 
By combining these two novel components - dual embeddings and dual attention mechanism - we develop a new \emph{Dual Attentive Sequential Learning (DASL)} model that bidirectionally transfers user preferences between domain pairs and thus provides more dynamic and effective cross domain recommendations. 

As a part of this work, we conduct extensive offline experiments on three real-world datasets and show that the proposed DASL model significantly and consistently outperforms all the state-of-the-art baseline models in terms of the CTR-based cross domain recommendation performance. We also conduct online A/B test at a major video streaming platform Alibaba-Youku and show that the DASL model outperforms the current production system by 7.07\% in the total number of video views.

In this paper, we make the following research contributions:

(1) We propose novel dual embedding and dual attention techniques and a method of incorporating them into the cross-domain recommendation model. We show in the paper that these two new methods are effective for providing cross domain sequential recommendations that progressively improve CTRs through the iterative training process.

(2) We develop the Dual Attentive Sequential Learning (DASL) model that bidirectionally transfers user preferences and user-item interactions between domain pairs by deploying the proposed dual embeddings and dual attention mechanisms.

(3) We conduct extensive offline and online experiments to demonstrate superiority of the proposed DASL model, which consistently and significantly outperforms all the baseline models, as well as the latest production system at a major video streaming platform Alibaba-Youku. 

The rest of the paper is organized as follows. We discuss the related work in Section 2 and present our proposed DASL model for providing sequential cross domain recommendations in Section 3. Experimental design on three real-world datasets is described in Section 4. The results and discussions are presented in Section 5. We also present the online A/B experiments in Section 6. Finally, Section 7 summarizes our contributions and concludes the paper.

\section{Related Work}
In this section, we summarize the related work of our proposed model as three categories: sequential recommendations, cross domain recommendations and dual learning methods.

\subsection{Sequential Recommendations}
Our work is related to deep-learning based sequential recommender system \cite{wang2019sequential} which suggests items that may be of interest to a user by modelling the sequential dependencies over the user-item interactions and capturing the current and recent preferences of the user. Popular sequential recommendation models utilize sequential neural network structures, including GRU \cite{hidasi2015session}, LSTM \cite{zhu2017next}, and BERT \cite{sun2019bert4rec} for extracting latent user and item representations as well as their high-order interactions and achieving satisfying recommendation performance. Some sequential recommendation models also take into account personalized and session-based information during the recommendation process, including DIN \cite{zhou2018deep}, DSIN \cite{feng2019deep}, DIEN \cite{zhou2019deep}, DeepFM \cite{guo2017deepfm}, Wide \& Deep \cite{cheng2016wide}, PNN \cite{qu2016product} and PURS \cite{li2020purs}, as different people would have different preferences towards the recommended items under different circumstances. In this paper, we inherit the fundamental idea of sequential recommendations to construct our novel cross-domain CTR prediction model.

\subsection{Cross Domain Recommendations}
Cross domain recommender system \cite{fernandez2012cross} is a powerful method to handle the cold-start and data sparsity problem commonly encountered in recommendation tasks. By leveraging auxiliary information from other domains, cross domain recommendation models are extended from single-domain recommendation settings based on the assumption that different behavioral patterns jointly characterize the way users interact with items of a certain domain \cite{sahebi2017cross,singh2008relational,hu2013personalized,loni2014cross}. Researchers have also proposed to use transfer learning techniques \cite{pan2010survey} to transfer user preferences from the target domain to the source domain for providing enhanced recommendations, including CoNet \cite{hu2018conet}, CCCFNet \cite{lian2017cccfnet} and so on. As CTR prediction task become increasingly important for online commerce platforms, some recent studies aim for providing cross domain sequential recommendations \cite{ma2019pi,ouyang2020minet} and have achieved great success in addressing the cold-start and data sparsity problem. While most of the existing studies focus on unidirectional user preference transfer from the source domain to the target domain, in this paper we propose a novel dual learning-based cross domain sequential recommender system that greatly improves its business performance.

\subsection{Dual Learning}
As we previously discussed, transfer learning \cite{pan2010survey} constitute an important method for constructing cross domain recommendation models, which identifies and incorporates the common knowledge structure that defines the domain relatedness during the learning process. Several cross domain recommendation models \cite{zhuang2010collaborative,liu2013multi,long2013transfer} have successfully applied the idea of transfer learning to achieve satisfying recommendation performance. As the dual learning mechanism \cite{long2012dual,zhong2009cross,wang2011cross,he2016dual} is capable of exploiting the duality between two recommendation tasks and enhancing the transfer learning capability, researchers has purposed to utilize dual learning for providing cross domain recommendations \cite{li2020ddtcdr,zhu2019dtcdr,tkde2021} In particular, these models assume that if two users have similar preferences in a certain domain, their preferences should also be similar across other domains as well. While these dual learning-based cross domain recommendation models achieves significantly better performance over the classical models, they focus primarily on the static rating predict task, without taking into account user sequential behavior and the next-item recommendation task, which is the main focus of this paper.

\section{Model}
\begin{figure*}
\centering
\includegraphics[width=\textwidth]{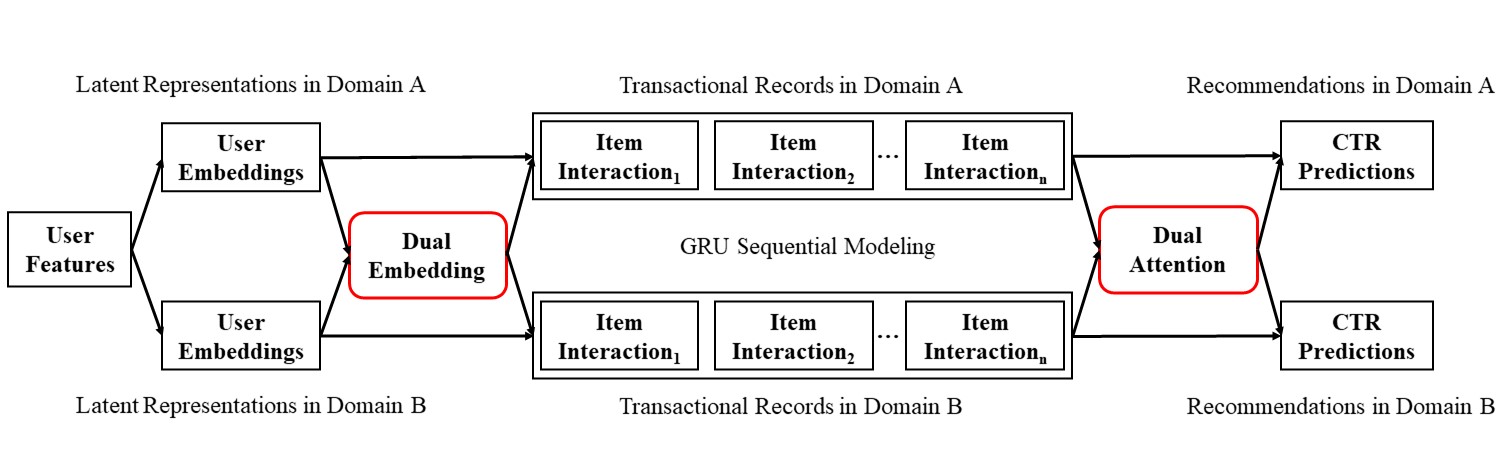}
\caption{Overview of the proposed DASL framework for providing cross domain sequential recommendations.}
\label{model}
\end{figure*}

In this section, we present the proposed Dual Attentive Sequential Learning (DASL) model for providing cross domain sequential recommendations, which is shown in Figure \ref{model}. First, we construct latent representations of users and items in both domains using the autoencoding technique. Then we utilize the GRU network that takes sequences of user consumptions as inputs and extracts user preferences accordingly. During this process, we design a novel Dual Embedding component that maps user representations into a shared latent space through latent orthogonal mapping functions. By doing so, we bidirectionally transfer latent user preferences across different domains through minimization of metric distances in the shared space. The final recommendation is produced through the matching process between user preferences and the candidate items, where we  utilize a novel Dual Attention component that takes into account relative importance of each consumption record in both domains. We will explain the details of our design in the following sections.

\subsection{User and Item Representations}
To effectively extract latent user preferences and efficiently model the interrelations of explicit features of users and items, we utilize the common practices of autoencoding techniques that automatically transforms the heterogeneous and discrete feature vectors into continuous feature embeddings. We denote the explicit features of user $a$ as $u_{a} = \{u_{a_{1}},eu_{a_{2}},\cdots,u_{a_{m}}\}$ and the explicit features of item $b$ as $i_{b} = \{i_{b_{1}},i_{b_{2}},\cdots,i_{b_{n}}\}$. The autoencoder framework trains two separate neural networks simultaneously: the encoder network, which maps explicit features into latent embeddings; and the decoder network, which reconstructs feature vectors from latent embeddings. Due to the effectiveness and efficiency of the Multi-Layer Percetron (MLP) network, we formulate both the encoder and the decoder as MLP, which learns the hidden representations by optimization the reconstruction loss $L$ for users and items as $L_{u} = ||u_{a}-MLP_{dec}(MLP_{enc}(u_{a}))||$ and $L_{i} = ||i_{b}-MLP_{dec}(MLP_{enc}(i_{b}))||$,where $MLP_{enc}$ and $MLP_{dec}$ represent the MLP networks for the encoder and the decoder respectively. Note that, during the learning process we construct different autoencoder networks for users and items in different domains to avoid information leakage between the domains.

\subsection{Dual Embedding}
While the user representations in different domains are obtained through the same autoencoding framework described in the previous section, the distribution of those latent embeddings would still be vastly different, as the user behaviors and preferences are highly heterogeneous across different domains. To utilize heterogeneous user behavioral information and address such phenomenon, we identify and extract overlap users that have interacted with items in both domains and use them as `pivots' to learn the relations of user preferences and behaviors in different domains.

In this section, we present the novel Dual Embedding component, which is constructed by identifying users with similar preferences in one domain and bidirectionally transfer this similarity information to the other domain. In particular, we adopt the idea of metric learning \cite{kulis2012metric} and map the latent embeddings in different domains into a shared latent space through the transitional mapping. By minimizing the metric distances in the shared latent space, this transitional mapping would pull the mapped embeddings of the same user from different domains closer, while pushing embeddings from different users further apart. According to the triangle inequality, this metric learning process will also pull those users who purchase similar items across different domains together. Eventually, the nearest neighbor points for any given user in the shared latent space will become its representation in the other domains respectively. Through the metric learning mapping that effectively captures the relations between overlap users, we propagate the learning process that not only models the information of overlap users, but also non-overlap users and user-item interactions, which is typically difficult to capture in cross-domain recommendation tasks.

Different from the classical metric learning settings, we restrict this transitional mapping to be an orthogonal mapping to obtain certain benefits for the model optimization process, following the discussion in \cite{li2020ddtcdr}:
\begin{itemize}
\item Orthogonality preserves similarity between embeddings of different users during the metric learning process since orthogonal transformation preserves inner product of latent vectors. 
\item Orthogonal mapping $X$ has transpose mapping $X^T$ as its inverse \cite{greub2012linear}, which simplifies the learning procedure and enables efficient dual learning process. 
\item Orthogonality restricts the parameter space of the transitional mapping and thus reduces the computational complexity of the learning process.
\end{itemize}

Specifically, we denote the user embeddings for domain $A$ and $B$ as $W_{u_{A}}$, $W_{u_{B}}$ that we obtained in the previous section. The optimization goal is to search for the mapping matrix $X$ that minimizes the sum of squared Euclidean distances between the mapped user embeddings $XW_{u_{A}}$ and the target user embeddings $W_{u_{B}}$ for the same overlap users: 
\begin{equation}
L_{X} = argmin_{X} \sum_{W_{u_{A}},W_{u_{B}}\in u_{A},u_{B}} ||XW_{u_{A}}-W_{u_{B}}||^{2}
\end{equation}
This optimization is equivalent to its dual form:
\begin{equation}
L_{X^{T}} = argmin_{X} \sum_{W_{u_{A}},W_{u_{B}}\in u_{A},u_{B}} ||W_{u_{A}}-X^{T}W_{u_{B}}||^{2}
\end{equation}
We constrain the metric learning mapping $X$ to be an orthogonal mapping (i.e. $XX^{T}=X^{T}X=I$), which serves to enforce structural invariance of user preferences in each domain, while preventing a degradation in mono-domain recommendation performance for learning better transitional mappings.

As we simultaneously map user embeddings from different domains into the shared latent space, we effectively utilize the dual learning mechanism that updates user embeddings for both domains using equations (1) and (2) simultaneously. The learning process can then be repeated iteratively to obtain progressively better cross-domain recommendation performance until the convergence criterion is met. As a result, it pushes the metric learning framework to better capture user preferences and thus provide even better recommendation performance.

\subsection{Sequential Learning}
Next, we construct latent representations of user preferences from their past consumptions that account for the personalized, dynamic and contextual factors during the recommendation process. Since user behavior streams can be modeled as temporal sequences, we adopt the state-of-the-art deep sequential learning models to condense user behavior streams into latent representations of user preferences to facilitate the learning process. In particular, we select the recurrent neural network to model user interests, for it is capable of capturing the time information and the sequence of user transactions, as more recent interactions would naturally have a higher impact on the current recommendation than previous interactions. As the GRU \cite{cho2014learning} neural network is shown to be computationally more efficient and is capable of providing better performance \cite{chung2014empirical}, we select GRU as our choice of the RNN network.

During the learning process, we first map the behavior sequence to the corresponding item embeddings obtained in the previous stage. To illustrate the GRU learning procedure, we denote $W_{z}$,$W_{r}$,$U_{z}$ and $U_{r}$ as the weight matrices of current information and the past information for the update gate and the reset gate respectively. $x_{t}$ is the user state input at timestep $t$, while $h_{t}$ stands for the user state output. $z_{t}$ denotes the update gate status and $r_{t}$ represents the status of reset gate. Therefore, the hidden state at timestep $t$ could be obtained following these equations:
\begin{equation}
z_t = \sigma_g(W_{z} x_t + U_{z} h_{t-1} + b_z) 
\end{equation}
\begin{equation}
r_t = \sigma_g(W_{r} x_t + U_{r} h_{t-1} + b_r) 
\end{equation}
\begin{equation}
h_t =  (1-z_t) \circ h_{t-1} + z_t \circ \sigma_h(W_{h} x_t + U_{h} (r_t \circ h_{t-1}) + b_h)
\end{equation}

By iteratively calculating hidden states throughout user behavior sequences, we would obtain the final hidden state at the end of user behavior sequences, which constitutes the representation of user preferences for the current recommendation. The latent representations of user preferences would be subsequently used for providing cross-domain recommendations.

\subsection{Dual Attention}
Note that each user past transaction might have different effect on the current recommendation depending on its time of occurrence and other contexts. As historical records of user transactions may contain items that are not directly relevant to the current recommendation choices, attention mechanism \citep{bahdanau2014neural} is often utilized to assign the transaction records with different weights to build an attentive context that outputs the proper next item with a high probability of a positive outcome. While previously proposed models have incorporated the self-attention mechanism \cite{shaw2018self,zhou2018deep} during the sequence learning process, they only consider the case of providing single-domain recommendations. To provide effective cross-domain recommendations, in this paper we propose a novel Dual Attention component, which aggregates user behaviors in \emph{both} domains that jointly determine the relevance of each consumption record towards the current item recommendation.

The Dual Attention component is constructed based on the framework of scaled dot-product attention \cite{vaswani2017attention}, which is much faster and more space-efficient in practice, compared to the additive attention framework \citep{bahdanau2014neural}. Unlike classical attention-based recommendation models that only takes into account user behavior in one single domain, we propose to incorporate user behaviors in multiple different domains simultaneously to obtain the associated attention values. By doing so, we could provide a more comprehensive and effective modeling of the matching process between user preferences and candidate items, as the relative importance of each historical interaction towards current recommendations would be affected by user interactions in the other domains as well. For example, if we know that a user has watched videos related to Harry Potter, we might be willing to recommend another Harry Potter video to that user, although we are not entirely certain about the user's preferences as the user has also watched other types of videos. However, if we also know that this user has finished reading the entire series of Harry Potter novels from her/his activities in the Book domain, we would assign a much higher weight to that user's Harry Potter video watching experiences in the Video domain, thus making it more likely and with higher confidence to recommend similar content to that user. Similarly, user information in the Video domain would also help us to estimate better the attention values in the Book domain.

To illustrate the learning process of the Dual Attention component, we denote the input of queries in both domains  as $Q_{1}$, $Q_{2}$ consisting past consumption records of the user in both domains. We also denote keys as $K$ of dimension $d_k$, and values as $V$ of dimension $d_v$, which will be updated during the learning process. We compute the dot products of both queries with all keys, divide each by $\sqrt{d_k}$ to counteract the effect that the dot products grow large in magnitude for large values of $d_k$. We subsequently apply a softmax function to obtain the attention weights on these values for both domains simultaneously as:

\begin{equation}
   \mathrm{Attention}(Q_{1}, K, V) = \mathrm{softmax}(\frac{(Q_{1};Q_{2})K^T}{\sqrt{d_k}})V
\end{equation}

\begin{equation}
   \mathrm{Attention}(Q_{2}, K, V) = \mathrm{softmax}(\frac{(Q_{2};Q_{1})K^T}{\sqrt{d_k}})V
\end{equation}

The final recommendations of our proposed Dual Attentive Sequential Learning model are obtained by aggregating the user embeddings generated through the Dual Embedding component described in Section 3.2, the latent representations of user preferences described in Section 3.3, and the attention values representing the relative importance of each consumption record obtained in Section 3.4. We feed the aggregated vectors into the Multi-Layer Perceprton (MLP) network to obtain the predicted values of the utility function. We produce the final recommendations by selecting the top-n items with the highest utility values. To demonstrate the superiority of our proposed model, we implement it on several industrial recommendation applications, which will be described in the next section.

\section{Experiments}
\subsection{Dataset}
We conduct extensive offline evaluations on three industrial cross-domain recommendation datasets to evaluate the performance of our proposed model. The Imhonet dataset \cite{DBLP:conf/cla/BobrikovNI16} contains user ratings and feedback information over the period of 2011 to 2014 across multiple domains. We select the two largest domains in the Imhonet, namely Books and Movies, to conduct our experiments. The Amazon dataset \cite{ni2019justifying} consists of user purchase actions and rating information collected from the Amazon platform, and we select two domains with sufficient amount of overlap users for conducting our experiments: Toys and Video Games. Finally, we test our model on the Youku dataset, which contains user logs for the domains of TV Shows and Short Videos of a major video-streaming platform Alibaba-Youku in China during the month of December 2020.

We list the basic statistics and the number of overlap users of these three datasets in Tables \ref{imhonet}, \ref{amazon} and \ref{imhonet} respectively. During the recommendation process, we identify the overlap users across different domains using a common user ID. We binarize the rating information as labels of click and non-click for the click-through-rate prediction task. Note that while each user might purchase items from different domains, each item belongs only to a single domain in all the three datasets. For providing sequential recommendations, we construct the purchasing history for each user as the last 10 purchased items based on the recorded timestamp for all experimental settings.

\begin{table}
\centering
\begin{tabular}{|l l l|}
\hline
Domain & Book & Movie \\ \hline
\#Users & 804,285 & 959,502 \\ \hline
\#Items & 182,653 & 79,866 \\ \hline
\#Records & 223,007,805 & 51,269,130 \\ \hline
Sparsity & 0.0157\% & 0.0669\% \\ \hline
Overlap Users & \multicolumn{2}{c|}{318,225}\\ \hline
\end{tabular}
\caption{Descriptive Statistics of the Imhonet Dataset}
\label{imhonet}
\end{table}

\begin{table}
\centering
\begin{tabular}{|l l l|}
\hline
Domain & Toys & Video Games \\ \hline
\#Users & 1,342,911 & 826,767 \\ \hline 
\#Items & 327,698 & 50,210 \\ \hline
\#Records & 2,252,771 & 1,324,753 \\ \hline
Sparsity & 0.0005\% & 0.0032\% \\ \hline
Overlap Users & \multicolumn{2}{c|}{131,684}\\ \hline
\end{tabular}
\caption{Descriptive Statistics of the Amazon Dataset}
\label{amazon}
\end{table}

\begin{table}
\centering
\begin{tabular}{|l l l|}
\hline
Domain & TV Shows & Short Videos \\ \hline
\#Users & 63,360 & 63,360 \\ \hline
\#Items & 17,966 & 941,507 \\ \hline
\#Records & 11,558,124 & 19,162,111 \\ \hline
Sparsity & 1.0150\% &  0.0321\% \\ \hline
Overlap Users & \multicolumn{2}{c|}{63,360}\\ \hline
\end{tabular}
\caption{Descriptive Statistics of the Youku Dataset}
\label{industrial}
\end{table}

\subsection{Baseline Models and Evaluation Metrics}
To demonstrate the superiority of our proposed Dual Attentive Sequential Learning (DASL) model, we compare it with the following three categories of state-of-the-art baseline recommendation models. We also select two popular accuracy metrics for the evaluation process, namely \textbf{AUC} and \textbf{HR@10} \cite{gunawardana2015evaluating}. For all the experiments, we conduct 5-fold cross validation and report the average recommendation performance.

\subsubsection{Cross-Domain Sequential Recommendations}
\begin{itemize}
\item \textbf{Pi-Net \cite{ma2019pi}} The Parallel Information-sharing Network (Pi-Net) simultaneously generate recommendations for two domains through shared user accounts. It consists of two core units: a shared account filter unit to learn user-specific representation and a cross-domain transfer unit to transfer user information.
\item \textbf{MiNet \cite{ouyang2020minet}} The Mixed Interest Network (MiNet) jointly models three types of user interest: 1) long-term interest across domains, 2) short-term interest from the source domain and 3) short-term interest in the target domain to provide cross-domain recommendations.
\end{itemize}

\subsubsection{Cross-Domain Recommendations}
\begin{itemize}
\item \textbf{DDTCDR \cite{li2020ddtcdr}} Deep Dual Transfer Cross Domain Recommendation (DDTCDR) efficiently transfers user preferences across domain pairs through dual learning mechanism.
\item \textbf{CoNet \cite{hu2018conet}} Collaborative Cross Networks (CoNet) enables knowledge transfer process across domains through cross connections between base networks.
\end{itemize}

\subsubsection{Sequential Recommendations}
\begin{itemize}
\item \textbf{DIN \cite{zhou2018deep}} Deep Interest Network (DIN) designs a local activation unit to adaptively learn the representation of user interests from historical behaviors with respect to a certain item.

\item \textbf{Wide \& Deep \cite{cheng2016wide}} Wide \& Deep utilizes the wide model to handle the manually designed cross product features, and the deep model to extract nonlinear relations among features.
\end{itemize}

\section{Results}

\begin{figure}[h]
    \centering
    \begin{subfigure}[t]{0.5\textwidth}
        \centering
        \includegraphics[width=\textwidth]{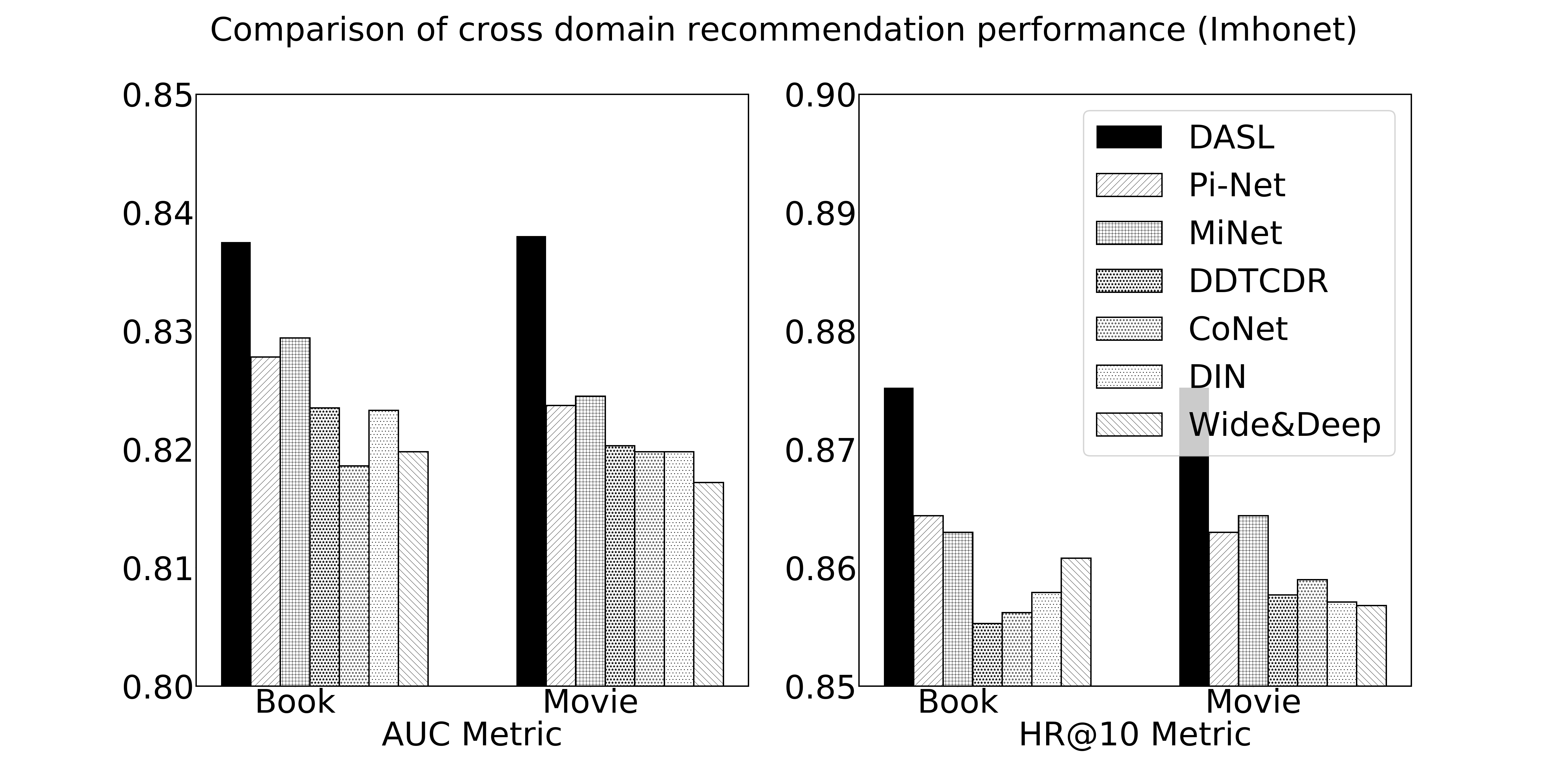}
        \caption{Imhonet Dataset}
    \end{subfigure}%
    \vskip\baselineskip
    \begin{subfigure}[t]{0.5\textwidth}
        \centering
        \includegraphics[width=\textwidth]{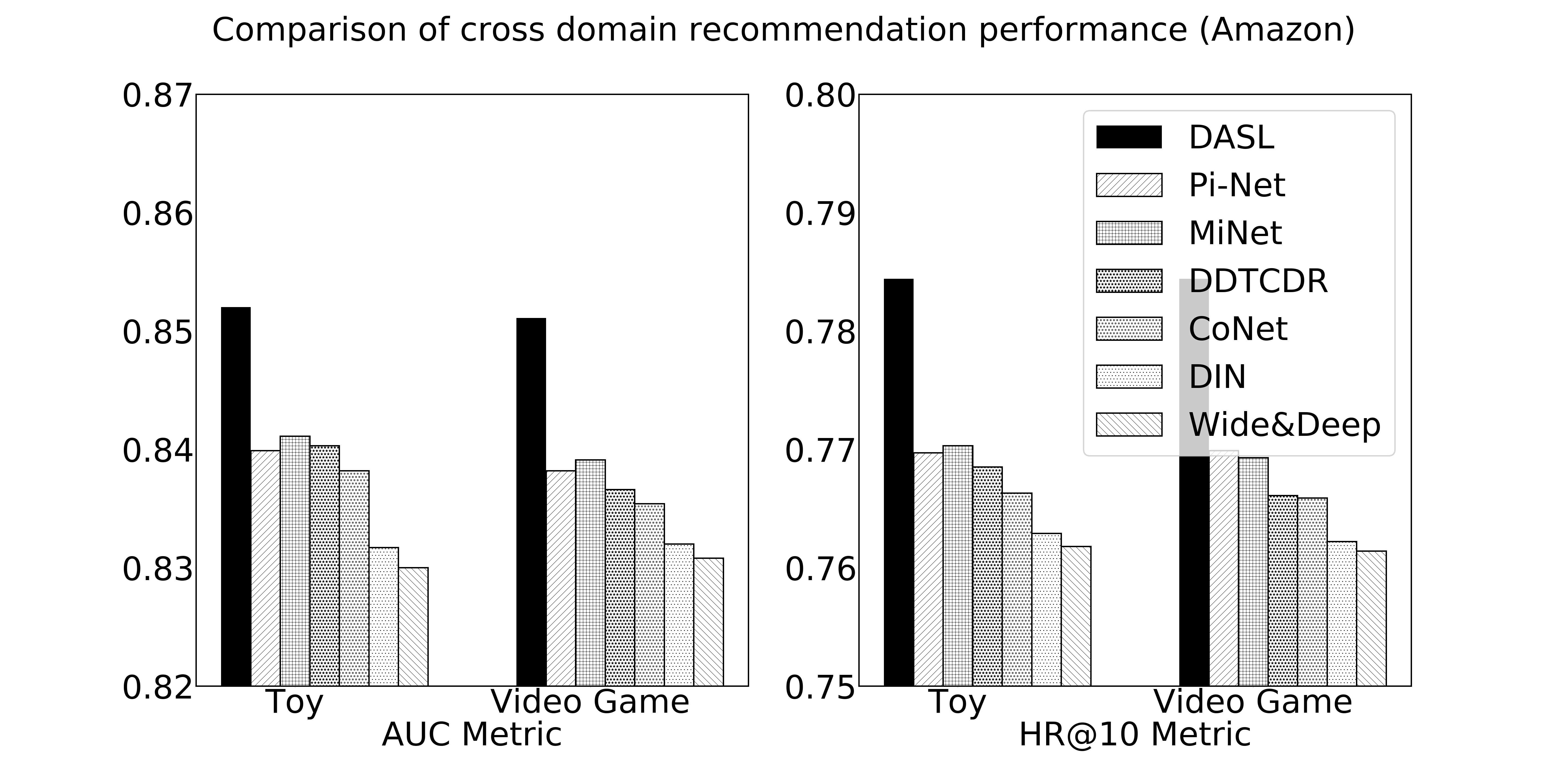}
        \caption{Amazon Dataset}
    \end{subfigure}%
   \vskip\baselineskip
    \begin{subfigure}[t]{0.5\textwidth}
        \centering
        \includegraphics[width=\textwidth]{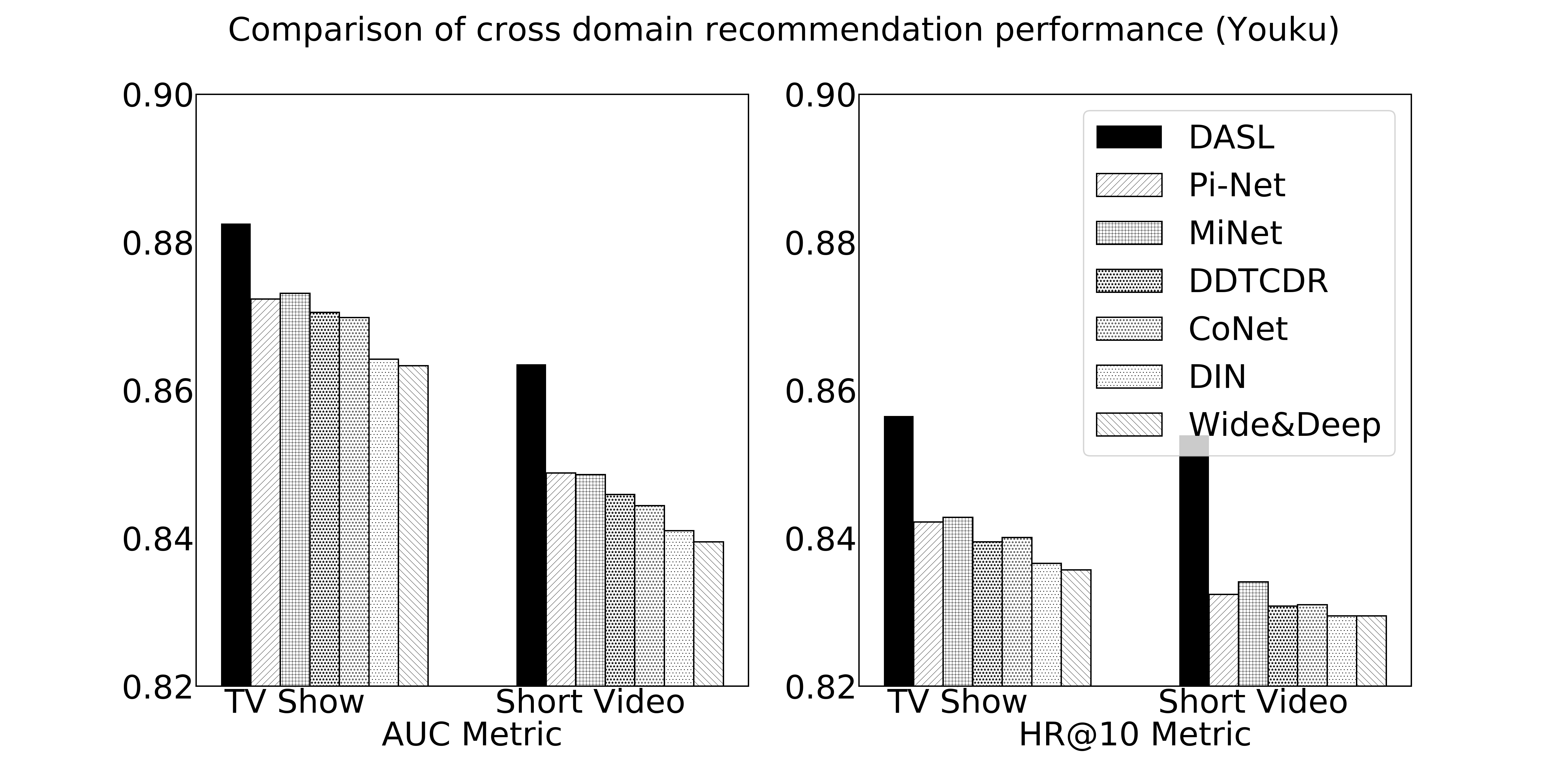}
        \caption{Youku Dataset}
    \end{subfigure}%
\caption{Comparison of cross domain recommendation performance in terms of AUC and HR@10 measures in three datasets.}
\label{comparison}
\end{figure}

\subsection{Cross-Domain Recommendation Performance}
The results of applying our proposed DASL model and the corresponding baseline models to the three industrial datasets under the experimental settings described in Section 4 are presented in Table \ref{result} and Figure \ref{comparison}. As Table \ref{result} and Figure \ref{comparison} illustrate, our DASL model significantly and consistently outperforms all other baselines in terms of the AUC and HR@10 metrics across all the three datasets. Most importantly, we improve the recommendation performance for both domains in every recommendation task, instead of only improving the performance of the target domain, as was typically done by the classical cross-domain recommendation models. In particular, we observe an average increase of 1.36\% for the AUC metric and 1.72\% for the HR@10 metric, compared to the second-best baseline approach. This clearly demonstrates the effectiveness of our dual learning design.

We also observe in our experimental settings that cross domain recommendation models (i.e., DASL, Pi-Net, MiNet, DDTCDR and CoNet) significantly outperform the single-domain recommendation models (i.e., DIN and Wide \& Deep). This is the case, as cross-domain recommendation models utilize user behavioral information from multiple domains, therefore obtaining more comprehensive understanding of user preferences for recommendation purposes, compared to single-domain recommendation models. In addition, we observe that cross domain sequential recommendation models (DASL, Pi-Net and MiNet) achieve significantly better recommendation results than static cross domain recommendation methods (DDTCDR and CoNet), which illustrates the importance of incorporating sequential user behavior during the recommendation process.

To summarize, all these results show that the dual learning mechanism is powerful for the cross domain sequential recommendation tasks and works well in practice, as the performance of our proposed DASL model attests to.

\begin{table*}
\centering
\resizebox{\textwidth}{!}{
\begin{tabular}{|c|cccc|cccc|cccc|} \hline
Algorithm & \multicolumn{4}{c|}{Imhonet} & \multicolumn{4}{c|}{Amazon} & \multicolumn{4}{c|}{Youku} \\ \cline{2-13}
               & \multicolumn{2}{c|}{Books} & \multicolumn{2}{c|}{Movies} & \multicolumn{2}{c|}{Toys} & \multicolumn{2}{c|}{Video Games} & \multicolumn{2}{c|}{TV Shows} & \multicolumn{2}{c|}{Short Videos} \\ \cline{2-13}
               & AUC & HR@10 & AUC & HR@10 & AUC & HR@10 & AUC & HR@10 & AUC & HR@10 & AUC & HR@10 \\ \hline
\textbf{DASL} & \textbf{0.8375*} & \textbf{0.8752*} & \textbf{0.8380*} & \textbf{0.8752*} & \textbf{0.8520*} & \textbf{0.7844*} & \textbf{0.8511*} & \textbf{0.7844*} & \textbf{0.8825*} & \textbf{0.8565*} & \textbf{0.8635*} & \textbf{0.8539*} \\
Increased \% & (+0.98\%) & (+1.25\%) & (+1.64\%) & (+1.25\%) & (+1.30\%) & (+1.83\%) & (+1.43\%) & (+1.96\%) & (+1.08\%) & (+1.63\%) & (+1.73\%) & (+2.37\%)\\ \hline
Pi-Net & 0.8278 & 0.8644 & 0.8237 & 0.8630 & 0.8399 & 0.7697 & 0.8382 & 0.7699 & 0.8723 & 0.8422 & 0.8488 & 0.8324 \\
MiNet & 0.8294 & 0.8630 & 0.8245 & 0.8644 & 0.8411 & 0.7703 & 0.8391 & 0.7693 & 0.8731 & 0.8428 & 0.8486 & 0.8341 \\ \hline
DDTCDR & 0.8235 & 0.8553 & 0.8203 & 0.8577 & 0.8403 & 0.7685 & 0.8366 & 0.7661 & 0.8705 & 0.8395 & 0.8459 & 0.8308 \\
CoNet & 0.8186 & 0.8562 & 0.8198 & 0.8590 & 0.8382 & 0.7663 & 0.8354 & 0.7659 & 0.8698 & 0.8401 & 0.8444 & 0.8310 \\ \hline
DIN & 0.8233 & 0.8579 & 0.8198 & 0.8571 & 0.8317 & 0.7629 & 0.8320 & 0.7622 & 0.8642 & 0.8366 & 0.8410 & 0.8295 \\
Wide \& Deep & 0.8198 & 0.8608 & 0.8172 & 0.8568 & 0.8300 & 0.7618 & 0.8308 & 0.7614 & 0.8633 & 0.8357 & 0.8395 & 0.8295 \\ \hline
\end{tabular}
}
\newline
\caption{Comparison of Cross-Domain Sequential Recommendation Performance}
\label{result}
\end{table*}

\subsection{Ablation Study}

\begin{table*}
\centering
\resizebox{\textwidth}{!}{
\begin{tabular}{|c|cccc|cccc|cccc|} \hline
Algorithm & \multicolumn{4}{c|}{Imhonet} & \multicolumn{4}{c|}{Amazon} & \multicolumn{4}{c|}{Youku} \\ \cline{2-13}
               & \multicolumn{2}{c|}{Books} & \multicolumn{2}{c|}{Movies} & \multicolumn{2}{c|}{Toys} & \multicolumn{2}{c|}{Video Games} & \multicolumn{2}{c|}{TV Shows} & \multicolumn{2}{c|}{Short Videos} \\ \cline{2-13}
               & AUC & HR@10 & AUC & HR@10 & AUC & HR@10 & AUC & HR@10 & AUC & HR@10 & AUC & HR@10 \\ \hline
\textbf{DASL} & \textbf{0.8375*} & \textbf{0.8752*} & \textbf{0.8380*} & \textbf{0.8752*} & \textbf{0.8520*} & \textbf{0.7844*} & \textbf{0.8511*} & \textbf{0.7844*} & \textbf{0.8825*} & \textbf{0.8565*} & \textbf{0.8635*} & \textbf{0.8539*} \\ \hline
DASL-DE & 0.8333 & 0.8704 & 0.8355 & 0.8714 & 0.8468 & 0.7803 & 0.8471 & 0.7799 & 0.8771 & 0.8523 & 0.8600 & 0.8498 \\
DASL-DA & 0.8341 & 0.8712 & 0.8358 & 0.8710 & 0.8482 & 0.7795 & 0.8474 & 0.7797 & 0.8777 & 0.8529 & 0.8600 & 0.8502 \\
Single Domain & 0.8234 & 0.8580 & 0.8197 & 0.8569 & 0.8319 & 0.7629 & 0.8320 & 0.7621 & 0.8642 & 0.8366 & 0.8412 & 0.8293 \\ \hline
\end{tabular}
}
\newline
\caption{The Ablation Study}
\label{ablation}
\end{table*}

As discussed in the previous section, our proposed DASL model achieves significant improvements over other baselines. These improvements indeed come from incorporating the two novel components into the design of recommendation model: \textbf{Dual Embedding}, which unifies the learning process of user representations in both domains, thus enabling progressive improvements of user preference modeling through an iterative learning process; and \textbf{Dual Attention}, which estimates the relative importance of historical items in both domains simultaneously for providing next-item recommendations. 

In this section, we conduct the ablation study to justify importance of each factor. Specifically, we compare the proposed model with the following variations:
\begin{itemize}
\item \textbf{DASL-DE} This is a variant of our proposed model, where we drop the Dual Embedding (DE) component and generate latent representation of explicit user features separately for each domain.
\item \textbf{DASL-DA} This is a variant of our proposed model, where we drop the Dual Attention (DA) component and obtain the attention values of user records separately for each domain.
\item \textbf{Single-Domain} This is a variant of our proposed model, where we drop both the Dual Embedding and the Dual Attention components during the recommendation process. That is to say, we train the recommendation model separately for different domains without transferring user preferences.
\end{itemize}

As shown in Table \ref{ablation}, if we remove any of these two novel components out of the recommendation model, we observe significant loss in both the AUC and HR@10 metrics. Therefore, the ablation study demonstrates that the superiority of our proposed model really comes from the combination of these novel features that all play significant role in the dual learning-based model and all contribute to superior cross-domain recommendation performance.

\section{Online A/B Test}
We have also conducted an online A/B test at a major video streaming platform Alibaba-Youku providing cross-domain recommendations of TV Shows and Short Videos in January 2021. During the testing period, we compared the proposed DASL model with the latest production model deployed in the company. We measured model performance using the standard business metric \textbf{VV} (Video View) extensively used in the company that specifies the average amount of videos or TV shows each user watches in one session. It is closely related to the Click-Through-Rate (CTR) metric that we optimize in our recommendation model. We find that our DASL model increased the VV metric in both the TV Shows and the Short Videos domains by 7.07\%, compared to the latest production system in the company, and this improvement is significant and consistent through our experimental process. Therefore, we demonstrate the superior performance of our DASL model and the power of incorporating the proposed Dual Embedding and the Dual Attention components into the cross domain recommendation toolbox. Since the proposed DASL model has achieved such strong performance results, it is currently in the process of being deployed at the company.

\section{Conclusions}
Previously proposed cross domain sequential recommendation models primarily leverage information from the source domain to improve CTR predictions in the target domain, without taking into account the duality nature of providing cross domain recommendations simultaneously for source-target domain pairs. In this paper, we propose to apply the dual learning mechanism to sequantial learning models to bidirectionally transfers user preferences across different domains simultaneously for cross-domain click-through rate prediction task in an iterative manner until the learning process stabilizes. In particular, the proposed Dual Attentive Sequential Learning (DASL) model consists of two novel components: the Dual Embedding component that simultaneously extracts user preferences in both domains, and the Dual Attention component that jointly determines the relative importance of each consumption record in both domains to produce final recommendations.

To demonstrate the superiority of our proposed model, we conduct extensive offline experiments on three real-world datasets and demonstrate that our proposed DASL model significantly and consistently outperformed several state-of-the-art baselines across all the experimental settings. We also conduct an online A/B test at a major video streaming platform, where our DASL model also significantly improves business performance results vis-a-vis the latest production system. Since the proposed model has achieved such strong performance results, it is currently in the process of being deployed at the company.

As the future work, we plan to extend the proposed components of Dual Embedding and Dual Attention to provide recommendations across multiple domains rather than only in domain pairs. We also plan to study theoretical properties, especially the convergence behavior, of our proposed framework.

\bibliographystyle{ACM-Reference-Format}
\balance
\bibliography{sigproc}

\end{document}